\documentclass[prd,aps,nofootinbib]{revtex4}
\usepackage{framed}
\usepackage{bbold}
\usepackage{slashed}
\usepackage{graphicx,hyperref}
\usepackage{amsmath}
\usepackage{amssymb}
\usepackage{epsfig}
\usepackage{hhline}
\usepackage{soul}
\usepackage{tabularx,pifont,multirow}
\usepackage{enumitem}
\usepackage{colordvi}
\usepackage{xcolor}
\usepackage{yfonts}
\usepackage{multido}

\newcommand{\be}{\begin{equation}}
\newcommand{\ee}{\end{equation}}

\newcommand{\CC}{\Lambda}
\newcommand{\rL}{\rho_{\Lambda}}

\newcommand{\MPl}{{\cal M}_{\rm Pl}}

\newcommand{\rvo}{\rho^0_{\rm vac}}
\newcommand{\rv}{\rho_{\rm vac}}
\newcommand{\rco}{\rho^0_{c}}
\newcommand{\rLb}{\rho_{\CC}^{({\rm b})}}

\newcommand{\wv}{w_{\rm vac}}

\newcommand{\Pv}{P_{\rm vac}}
\newcommand{\cH}{\mathcal{H}}
\newcommand{\nueff}{\nu_{\rm eff}}
\newcommand{\mpl}{m_{\rm Pl}}
\newcommand{\txi}{\tilde{\xi}}

\newcommand{\astar}{a_{*}}

\newcommand{\rI}{\rho_I}
\newcommand{\rmo}{\rho_{m}^0}

\makeatletter
\ams@newcommand{\vardot}[2]{%
  {\mathop{#2\kern0pt}\limits^{\vbox to-1.4\ex@{\kern-\tw@\ex@
   \hbox{\normalfont\multido{}{#1}{.}}\vss}}}}
\makeatother

\definecolor{darkgreen}{rgb}{0,0.3,0.05}
\makeatletter                                                         %
\newcommand*\rel@kern[1]{\kern#1\dimexpr\macc@kerna}                  %
\newcommand*\widebar[1]{                                              %
  \begingroup                                                         %
  \def\mathaccent##1##2{                                              %
    \rel@kern{0.8}                                                    %
    \overline{\rel@kern{-0.8}\macc@nucleus\rel@kern{0.2}}             %
    \rel@kern{-0.2}                                                   %
  }                                                                   %
  \macc@depth\@ne                                                     %
  \let\math@bgroup\@empty \let\math@egroup\macc@set@skewchar          %
  \mathsurround\z@ \frozen@everymath{\mathgroup\macc@group\relax}     %
  \macc@set@skewchar\relax                                            %
  \let\mathaccentV\macc@nested@a                                      %
  \macc@nested@a\relax111{#1}                                         %
  \endgroup                                                           %
}                                                                     %
\makeatother                                                          %

\begin{document}

%

\title{\LARGE {\bf  Equation of state of the running vacuum} \vspace{0.0cm}}

\author{\large \bf Cristian Moreno-Pulido\footnote{cristian.moreno@fqa.ub.edu} and Joan Sol\`a Peracaula\footnote{sola@fqa.ub.edu}\vspace{0.5cm}}

\affiliation{Departament de F\'isica Qu\`antica i Astrof\'isica, \\
and \\ Institute of Cosmos Sciences,\\ Universitat de Barcelona, \\
Av. Diagonal 647, E-08028 Barcelona, Catalonia, Spain}

\vskip0.5cm


\begin{abstract}
\vspace{0.05cm}
Recent studies of quantum field theory in FLRW spacetime suggest that the cause of the speeding up of the universe is  the running vacuum (RV), see \cite{CristianJoan2022,CristianJoan2020}.  Appropriate renormalization of the energy-momentum tensor shows that the vacuum energy density  is a smooth function of the Hubble rate and its derivatives: $\rv=\rv(H, \dot{H},\ddot{H},...)$.  This is because in QFT  the quantum scaling of $\rv$ with the renormalization point turns into  cosmic evolution with $H$.  As a result, any two nearby points of the cosmic expansion during the standard FLRW epoch are smoothly related through  $\delta\rv\sim {\cal O}(H^2)$.   In our approach, what we  call the `cosmological constant'  $\Lambda$  is just the nearly sustained value of  $8\pi G(H)\rv(H)$  around (any)  given epoch, where $G(H)$ is the running gravitational coupling.   In the present study, after summarizing the main QFT calculations supporting the RV approach,  we focus on  the calculation  of the  equation of state (EoS) of the RV for the entire cosmic history within such a QFT framework.  In particular, in the  very early universe, where higher (even) powers $\rv\sim{\cal O}(H^N)$  ($N=4,6,\dots$) triggered  inflation during a short period in which $H=$const, the  vacuum EoS  is very close to  $\wv=-1$. This ceases to be true during the  FLRW era,  where it  adopts the EoS of matter  during the relativistic ($\wv=1/3$) and non-relativistic ($\wv=0$) epochs. Interestingly enough, we find  that  in the late universe the EoS becomes mildly dynamical and mimics quintessence, $\wv\gtrsim-1$. It finally asymptotes to  $-1$ in the remote future, but in the transit the RV helps alleviating the $H_0$ and $\sigma_8$ tensions.
\end{abstract}
\maketitle

\section{Introduction: $\CC$ and  the cosmological constant problem}

After 105 years of history\,\cite{Einstein1917}, one of the most perplexing aspects of the cosmological constant (CC), $\CC$, in Einstein's gravitational field equations is that we still don't know what it is and why it has the value that we have measured.   It is usually associated to the energy density $\rvo=\CC/(8\pi G_N)$ of something that we call `vacuum' ($G_N$ being Newton's constant) and we call it vacuum energy density (VED). But we  don't know which vacuum we are referring to: is it  the cosmic vacuum,  or maybe the quantum mechanical vacuum, or else?  In addition, we naively assume that it remains strictly constant throughout the cosmic evolution. There is actually no need for that, since a (direct and/or indirect) dependence on the cosmic time, i.e. $\rv(t,\zeta)$,  is perfectly compatible with the Cosmological Principle, where $\zeta=\zeta(t)$ is some dynamical variable.  Still, we  prefer to believe that $\CC$ is a fundamental constant of Nature, maybe  because we feel that in this way  Occam's razor is safely on our side.  But soon we come across  a really nasty surprise: measurements show that its current value  is of order $\rvo\sim 10^{-47}$ GeV$^4\sim \left(10^{-3} {\rm eV}\right)^4$\,\cite{SNIa,PlanckCollab}  in natural units. Such a value turns out to be far too smaller  than any typical energy density in particle physics or quantum field theory (QFT), and hence we have not the slightest chance to provide a fundamental explanation for it.  We  realize that we are up against an unsurmountable brick wall:  the `cosmological constant problem'  (CCP),   which smashes Occam's razor  to pieces  in our hands, and with it all our hopes for a possible understanding of the universe on fundamental grounds.  The CCP  is indeed the baffling  realization that the successful QFT methods applied to the world of the elementary particles seem to predict an effective value for $\rv$ which is excruciatingly much larger than  the current critical density of the universe $\rco$  (which $\rvo$  should be comparable to)\,\cite{Weinberg89,CCPmore,JSPRev2013,JoanSolaPhilTransc2022}.
The  Higgs boson, whose discovery   (with  a mass $M_h\simeq 125$ GeV) was made just 10 years ago  certified the existence of the
electroweak vacuum from spontaneous symmetry breaking (SSB)\,\cite{BEH64}.  However, it presumably contributes  a huge (positive) amount $M_h^4\sim 10^8$ GeV$^4$   to the zero-point energy (ZPE) of the quantum vacuum, and also as much as  $\langle V\rangle\propto- M_h^2\,v^2\sim - 10^9$ GeV$^4$ (negative) from SSB,  with $v\sim 250$ GeV the Higgs vacuum expectation value (VEV).  No less  significant is
the ZPE part from the top quark, which is   $\propto -m_t^4\sim -10^9$ GeV$^4$  (negative because it is a
fermion).  With no a priori correlation between  ZPE and SSB,  we expect that our QFT estimates are wrong  by a factor of  $\left(10^9/10^{-47}\right)\sim 10^{56}$.  Yet,  this blatant fiasco pales when compared to the VED yield  from  quantum gravity: $\MPl^4/\rvo\sim 10^{120}$,  where $\MPl=\left(8\pi G_N\right)^{-1/2}=2.43 \times 10^{18}$~GeV is the (reduced) Planck mass. In the face of it, we are left flabbergasted and  impotent!

In the next section, we fly over some of  the troublesome issues that the notion of vacuum energy and  cosmological constant faces in the context of flat spacetime. A proper treatment is only possible in curved spacetime, and this is what the rest of the paper is about.

\section{Vacuum energy in flat spacetime}\label{sec:VEDMinkowski}

Because of the  CCP, the quantum vacuum option  for explaining dark energy (DE) with a $\CC$-term  became  outcast and was blamed of all evils, particularly of the acute fine tuning problem.  This is a bit unfair, of course,  as all existing  forms of DE are actually  plagued with the same tuning  illness and to a degree which is no lesser than that  of the quantum vacuum itself\,\cite{JSPRev2013,JoanSolaPhilTransc2022}.  Moreover, the  vacuum is a most fundamental notion in QFT;  we should expect that a description of the CCP and of the DE from first principles should actually come from the quantum vacuum and the machinery of QFT. A simple  calculation on renormalizing the VED of a single free scalar field $\phi$  in Minkowski spacetime, e.g. using  Minimal Subtraction (MS) and dimensional regularization (DR),  renders the following, well-known, one-loop result (see e.g. \cite{JSPRev2013,JoanSolaPhilTransc2022} and references therein):
\begin{equation}\label{VEDMink}
\rv=\rL(\mu)+\frac{m^4}{64\pi^2}\,\left(\ln\frac{m^2}{\mu^2}+C_{\rm
vac} \right)\,.
\end{equation}
Here $\rL(\mu)$ is the renormalized cosmological term in the Einstein-Hilbert (EH) action and $\mu$ is the usual 't Hooft's mass unit of DR\,\cite{Collins84}.  The second term on the \textit{r.h.s} is the MS-renormalized ZPE at one-loop.   In a symbolic way, we may write $ {\rm VED}=\rL+{\rm ZPE}$. This expression was made finite by the usual counterterm procedure:  $\rLb=\rL(\mu)+\delta\rL$, wherein $\rLb$ is the starting (bare) coupling in the EH action and $\delta\rL$ is the MS-counterterm in any of its variants, which leaves an arbitrary constant  $C_{\rm vac}$ in the result after cancelling a pole  in $n=4$ spacetime dimensions.  This is prima facie all very simple in Minkowski space, but simplicity is not at all an advantage here, for  Eq.\,\eqref{VEDMink} carries already  the whole drama of the CCP. If that expression is  interpreted as the VED, the ZPE part is proportional to $m^4$, and hence for any typical mass in particle physics we have to fine tune $\rL(\mu)$ to an incommensurable level (from $55$ to $120$ decimal places, see above) to produce $\rL+{\rm ZPE}\sim 10^{-47}$ GeV$^4$.  Not to mention the mandatory (hyperfine) retuning to be made at higher and higher orders of perturbation theory\,\cite{JSPRev2013}.

It is important  not to confuse VED with CC.  The former may exist in Minkowski spacetime, as given e.g. in Eq.\,\eqref{VEDMink}, whereas the latter can only exist in the context of Einstein's equations of curved spacetime and hence in the presence of gravity.  Only in the last case the CC is  physically meaningful  and its value becomes inexorably intertwined with the VED through Einstein's equations, as follows: $\rv=\CC/(8\pi G_N)$.  We should not confuse the physical $\CC$ defined in this way with the corresponding bare parameter in the EH-action, which is  related to $\rL$  in a similar way (but in this case the relation involves only the bare values of all the parameters involved).
The problem with the above calculation is that it is of no use at all in curved spacetime, say in the cosmological Friedman-Lemaitre-Robertson-Walker (FLRW) background. There is no sense in associating the scale $\mu$ to any cosmological variable since, if  Einstein's equations are invoked,  the $\CC$ term as such in these equations cannot exist in Minkowski space unless the VED is exactly $\rL+{\rm ZPE}=0$.  So there  is no cosmology to do with  Eq.\eqref{VEDMink}, despite some  attempts in the literature.  This point has  been driven home recently in \cite{JoanSolaPhilTransc2022},  and in  \cite{Mottola2022}.  A realistic approach to the VED  within QFT in curved spacetime  should be different. A recent attempt  has  been put forward in the comprehensive work\,\cite{CristianJoan2022}, which further  extends that of \cite{CristianJoan2020} in providing a QFT formulation of the   RV framework, or  running vacuum model (RVM). See also  the review  \,\cite{JoanSolaPhilTransc2022} for a summarized account and a generous list of references   Here we shall adopt this same  approach  in order to investigate the  equation of state (EoS) of the quantum vacuum. As we shall see, it  does not reduce to just the traditional  result $\wv=-1$.  It turns out that in a QFT formulation the vacuum EoS  becomes dynamical and evolves as  a nontrivial function of the cosmic expansion,  $\wv=\wv(H,\dot{H},\ddot{H},...)$, where dots indicate differentiation with respect to cosmic time $t$, i.e. $\dot{(\textrm{ })}\equiv d(\textrm{ })/dt$.

\section{Computing the vacuum energy density in FLRW spacetime}\label{sec:VEDFLRW}

 Before we can face the computation of the EoS of the running vacuum  in curved spacetime, we need to compute the vacuum energy density (VED) and vacuum pressure.  This is sooner said than done, and  we should not presume that they are related in the simple way $\Pv=-\rv$, which is valid only in the classical theory without quantum matter fields.   In this section and the next, we summarize the approach and the main results presented at length in \,\cite{CristianJoan2022} insofar as concerns the calculation and renormalization of the VED in FLRW spacetime\,\footnote{We adopt the same conventions of\,\cite{CristianJoan2022}. See,  in particular,  Appendix \ref{sec:AppendixA} of that reference.}.  The reader mainly  interested on the phenomenological results may now wish to jump  directly to Sec.\,\ref{sec:CanonicalRVM} and skip some QFT technicalities.

 To simplify  the (usually arduous) computations in curved spacetime,  and also to minimize the number of parameters involved, we use just a single quantum matter scalar field $\phi$  with mass $m$,  nonminimally coupled to curvature  and without effective potential, hence with the action:
\begin{equation}\label{eq:Sphi}
  S[\phi]=-\int d^nx \sqrt{-g}\left(\frac{1}{2}g^{\mu \nu}\partial_{\mu} \phi \partial_{\nu} \phi+\frac{1}{2}(m^2+\xi R)\phi^2 \right)\,.
\end{equation}
Even with this relatively simple system, in which $\phi$ has no interactions with other fields nor with itself, QFT calculations become already quite cumbersome\,\cite{CristianJoan2022}.   The ZPE associated to $\phi$ is, of course,  an UV-divergent quantity.  Parameter $\xi$ in the action  is the non-minimal coupling of $\phi$   to gravity.  It is well-known that in the special case  $\xi=1/6$, the massless ($m=0$)  action is (locally) conformal invariant in $n=4$ spacetime dimensions.  Although $\xi$ is not necessary for the QFT renormalization of the above action at one-loop, it  is convenient to keep $\xi$ arbitrary.  In general, the presence of a nonminimal coupling is expected in a variety of contexts, e.g. in extended gravity theories\,\cite{Sotiriou2010,Capozziello2008,Capozziello2011,CapozzielloFaraoni2011}.  There is also a fermionic contribution to the VED, of course,  but it  is not necessary for the present considerations\,\cite{Samira2022}. In this work, therefore, we shall focus on the scalar contribution only.

First of all, we must compute the  ZPE of $\phi$  in FLRW spacetime.  However, in contrast to the previous section, rather than keeping on MS-renormalization to deal with the UV divergences also in the curved spacetime case  (which proves  inappropriate to deal with the CCP\,\cite{JSPRev2013,JoanSolaPhilTransc2022}), we adhere to adiabatic renormalization\,\cite{BirrellDavies82,ParkerToms09,Fulling89},  where physical quantities are organized in the so-called adiabatic orders,  although with a crucial nuance: we renormalize the energy-momentum-tensor (EMT) off-shell, meaning that we define its renormalized VEV  (associated to the fluctuations $\delta\phi$  of the fields)  as follows\,\cite{CristianJoan2022,CristianJoan2020}:
\begin{eqnarray}\label{EMTRenormalized}
\langle T_{\mu\nu}^{\delta \phi}\rangle_{\rm Ren}(M)&=&\langle T_{\mu\nu}^{\delta \phi}\rangle(m)-\langle T_{\mu\nu}^{\delta \phi}\rangle^{(0-4)}(M)\,.
\end{eqnarray}
The latter, as can be seen,  is obtained by performing an appropriate substraction  from its on-shell value  (i.e. the value defined  on the mass  $m$ of the quantized field), specifically we subtract the vacuum EMT value (i.e. its VEV)  computed at an arbitrary scale $M$.   The result is finite because we subtract adiabatic orders up to order $4$  (the only ones that can be divergent in $n=4$).  This is entirely different from MS since we subtract both UV-divergent and convergent parts at $M$. The renormalization point $M$ will be used later on as a renormalization group (RG)  tool  to explore the cosmic evolution at the expansion history time $H(t)$ by setting $M=H$.  But here is left arbitrary.  Let us note that the renormalized EMT must be related with the (renormalized)  effective action of vacuum, namely the action $W$  describing the vacuum fluctuations of the quantized matter fields of QFT in FLRW spacetime,\,\cite{BirrellDavies82,ParkerToms09,Fulling89}:
\begin{equation}\label{eq:DefW}
\langle T^{\delta\phi}_{\mu\nu}\rangle=-\frac{2}{\sqrt{-g}} \,\frac{\delta W}{\delta g^{\mu\nu}}\,.
\end{equation}
This relation offers us a precious opportunity for a nontrivial cross-check.  In fact, one can choose any pathway: we may either compute \eqref{EMTRenormalized} directly  by expanding the solution of the Klein-Gordon equation $(\Box-m^2-\xi R)\phi=0$  (satisfied by the quantum field operator $\phi$ in FLRW spacetime) in Fourier  field modes and letting the creation and annihilation  operators to act on the vacuum with the usual commutation relations; or, alternatively, we may compute the (renormalized)  effective action  $W$ through the DeWitt-Schwinger expansion\,\cite{BirrellDavies82,ParkerToms09,Fulling89} (upon carefully correcting their coefficients to account for the off-shell effects at the scale $M$), and then use Eq.\,\eqref{eq:DefW} to retrieve the renormalized EMT.   Let us note, in particular, that the Fourier field modes of the first method must be computed using the WKB expansion assuming the notion of adiabatic vacuum (which all of the annihilation operators must destroy)\,\cite{Bunch1980}.  The details of this lengthy calculation can be found in the comprehensive studies \,\cite{CristianJoan2022, CristianJoan2020},   see \cite{JoanSolaPhilTransc2022} for a summarized exposition. The important point is that the two pathways  must converge, and do indeed converge, exactly to the same result.  Once the renormalized EMT is accounted for  by any of these procedures, we must extract the renormalized VED out of it.  We perform the calculation in the conformally flat metric, $ds^2=a^2(\tau)\eta_{\mu\nu}dx^\mu dx^\nu$, where  $\eta_{\mu\nu}={\rm diag} (-1, +1, +1, +1)$ is the Minkowski metric  ($\tau$ being the conformal time and $a$ the scale factor of the FLRW line element).  Since the renormalized VEV of the EMT at the scale $M$ takes the form  $ \langle T_{\mu\nu}^{\rm vac}\rangle_{\rm Ren}(M)=-\rho_\Lambda (M) g_{\mu \nu}+\langle T_{\mu \nu}^{\delta \phi}\rangle_{\rm Ren}(M)$, the renormalized VED  at that scale reads\,\footnote{The scale $M$ should not be confused with  't Hooft's mass unit $\mu$  in DR\,\cite{Collins84}. Both scales may appear simultaneously in the calculations, with $\mu$ playing here (optionally) a mere auxiliary role in  intermediate steps  (e.g. if one opts for using DR to deal with the divergent  integrals), see \cite{CristianJoan2022,CristianJoan2020}. Since, however, we are not using at all the MS scheme as a renormalization procedure, the renormalized results cannot depend on $\mu$, but only on $M$.  Needless to say, the full effective action does not depend on $M$ either, but the renormalized VED does since the effective action of vacuum is only a part of the full effective action\,\cite{CristianJoan2022}.}
\begin{equation}\label{RenVDE}
\rho_{\rm vac}(M)= \frac{\langle T_{00}^{\rm vac}\rangle_{\rm Ren}(M)}{a^2}=\rho_\Lambda (M)+\frac{\langle T_{00}^{\delta \phi}\rangle_{\rm Ren}(M)}{a^2}\,.
\end{equation}
We can see that the above expression also adopts the structure $ {\rm VED}=\rL+{\rm ZPE}$, where the  $00th$ component  (the ZPE)  emerges from the explicit calculation of  \eqref{EMTRenormalized} in the FLRW metric\,\cite {CristianJoan2022,CristianJoan2020}. Up to $4th$ adiabatic order,  a lengthy calculation yields the following compact result:
\begin{equation}\label{Renormalized2}
\begin{split}
&\langle T_{00}^{\delta \phi}\rangle^{(0-4)}_{\rm Ren}(M)
=\frac{a^2}{128\pi^2 }\left(-M^4+4m^2M^2-3m^4+2m^4 \ln \frac{m^2}{M^2}\right)\\
&-\left(\xi-\frac{1}{6}\right)\frac{3 \mathcal{H}^2 }{16 \pi^2 }\left(m^2-M^2-m^2\ln \frac{m^2}{M^2} \right)+\left(\xi-\frac{1}{6}\right)^2 \frac{9\left(2  \mathcal{H}^{\prime \prime} \mathcal{H}- \mathcal{H}^{\prime 2}- 3  \mathcal{H}^{4}\right)}{16\pi^2 a^2}\ln \frac{m^2}{M^2}\,.
\end{split}
\end{equation}
This expression is finite and explicitly dependent on the scale $M$, the mass $m$ of the particle and the conformal Hubble rate $\cH$  and its time derivatives in conformal time (related to the ordinary Hubble rate in cosmic time $t$ simply as $\cH(\eta)=a H(t)$). Primes denote differentiation with respect to conformal time:  $\left(\right)^\prime\equiv d\left(\right)/d\tau$.  With  $ \langle T_{\mu\nu}^{\rm vac}\rangle_{\rm Ren}(M)$ given as above, the renormalized vacuum part of the generalized  Einstein's equations within  QFT in curved spacetime read $\MPl^2 (M) G_{\mu \nu}+\alpha(M) ^{(1)}{\rm H}_{\mu \nu}= \langle T_{\mu\nu}^{\rm vac}\rangle_{\rm Ren}(M)$,  where $ ^{(1)}{\rm H}_{\mu \nu}$ is a standard higher derivative (HD) tensor\,\cite{BirrellDavies82}, the only one needed in conformally flat spacetimes (such as FLRW).
The VED   $\rv=\rv(M,H)$, as given by \eqref{RenVDE}-\eqref{Renormalized2},   is a function not only of $M$ but also an explicit function of  $H$ and of its time derivatives. The change of the VED with respect to $M$ and $H$  reads
\begin{equation}\label{DiffHH0MM0}
\begin{split}
\rv(M,H)-\rv(M_0,H_0)&
=\frac{3\left(\xi-\frac{1}{6}\right)}{16\pi^2}\left[H^2\left(M^2-m^2+m^2\ln\frac{m^2}{M^2}\right)\right.\\
&\left.-H_0^2\left(M_0^2-m^2+m^2\ln\frac{m^2}{M_0^2}\right)\right]+{\cal O}(H^4)\,.
\end{split}
\end{equation}
This result  is  a perfectly smooth function with no quartic mass terms  $\sim m^4$ (see next section).   In addition,  the quadratic  ones $\sim m^2$ become  completely smoothed by the $H^2$ factor. Whence,  the terms  $\sim m^2H^2$ are fully innocuous for the CCP; and, finally,  those of order ${\cal O}(H^4)$ are irrelevant for the current universe.
The above renormalization procedure of the VED is in accordance with the standard QFT formalism in curved spacetime, where the UV-divergences of the vacuum effective action $W$ -- defined in \eqref{eq:DefW} --   can be transferred to the couplings of the classical action, which can absorb the infinities into renormalized coupling constants\,\cite{BirrellDavies82}. These renormalized couplings depend of course on the renormalization scale $M$ as the vacuum action $W$ itself,  and with it the VED becomes also dependent on $M$. The dynamics of vacuum in the RVM  is then inherited  from setting $M=H$ in the renormalized theory.   This is akin to set the renormalization scale to the characteristic energy of the process in ordinary gauge theories, a usual practice in the renormalization group approach\,\cite{JoanSolaPhilTransc2022}. In cosmology we have less clues on how to proceed, but in FLRW spacetime that setting looks reasonable and moreover it can be tested, see Sec. \ref{sec:CanonicalRVM}.  We should emphasize, however,  that while the vacuum action $W$ does depend on $M$  the full effective action containing also the classical part  does not depend on the scale $M$. This is of course the essence of the renormalization group and thanks to this condition one can derive the renormalization group equations for all the couplings, see\,\cite{CristianJoan2022}.  A particularly important renormalization group equation is that of the VED itself and is discussed in the next section.

\section{$\beta$-function for the Vacuum Energy Density}\label{sec:betafunctionVED}

A chief result which can be derived  from equations \eqref{RenVDE} and \eqref{Renormalized2}  is the expression for the  $\beta$-function driving the RG-running of the vacuum energy density, $\rv$. This important  result was not know until very  recently \,\cite{CristianJoan2022}:
\begin{equation}\label{eq:RGEVED1}
\begin{split}
\beta_{\rv}=M\frac{\partial\rv(M)}{\partial M}=\left(\xi-\frac{1}{6}\right)\frac{3 {H}^2 }{8 \pi^2 }\left(M^2-m^2\right)
+\left(\xi-\frac{1}{6}\right)^2 \frac{9\left(\dot{H}^2 - 2 H\ddot{H} - 6 H^2 \dot{H} \right)}{8\pi^2}\,,
\end{split}
\end{equation}
where we have used   $\mathcal{H}^\prime=a^2(H^2+\dot{H})$ and
$\mathcal{H}^{\prime\prime}=a^3\left(2H^3+4 H\dot{H}+\ddot{H}\right)$ in \eqref{Renormalized2}, and also  the fact that the $\beta$-function for the renormalized parameter $\rL$ in the EH-action is
\begin{equation}\label{eq:betarhoLambda}
  \beta_{\rL} (M)=M \frac{\partial \rL(M)}{\partial M}=\frac{1}{2(4\pi)^2}(M^2-m^2)^2\,.
\end{equation}
The latter ensues from the fact that in Minkowski space ($H=0$) the expression \eqref{RenVDE} must be RG invariant, as it is indeed the case with \eqref{VEDMink} in the MS scheme\,\cite{JoanSolaPhilTransc2022}.  In  both renormalization schemes, the flat spacetime expressions correspond originally to renormalizing  a bare coupling and hence they are globally independent of $M$ (the renormalization point)\footnote{In Minkowski spacetime there is nothing else in the vacuum action apart from the term $\rL$.  In curved spacetime, in contrast, we have also the curvature  scalar  plus the geometric HD terms.  The renormalization of the VED is then not just the renormalization of a bare term, as in fact the VED becomes explicitly dependent on $H$  as well as on $M$, as we have just seen.  In this case,  only the full effective action (involving the classical part plus the nontrivial quantum vacuum effects) is scale- (i.e. RG-) independent, as previously noted. For a more formal derivation of these expressions using the full effective action, see \,\cite{CristianJoan2022}.}. Again the terms   ${\cal O}(H^4)$  are irrelevant for the present universe. The obtained $\beta$-function of the VED is thus very softly dependent on the mass scale, just  as  $\beta_{\rv}\propto {H}^2 \left(M^2-m^2\right) + {\cal O}(H^4)$ rather than the traditional (and troublesome!)  form  $\beta_{\rv}\propto m^4$.  This explains the cancellation of quartic terms in performing the subtraction \eqref{DiffHH0MM0} in the previous section\,\cite{CristianJoan2022}.  Thus, when one considers the evolution of the VED in this context there is no influence whatsoever from the dangerous quartic mass terms in\,\eqref{Renormalized2}.   It should be stressed  that the result \eqref{eq:RGEVED1} is exact and does not depend on the fact that \eqref{Renormalized2} was computed up to  $4th$ adiabatic order, as the higher order terms (order $6th$ and above)  are finite and hence do not depend on $M$.

\begin{figure}[t]
\begin{centering}
  \includegraphics[width=1\textwidth]{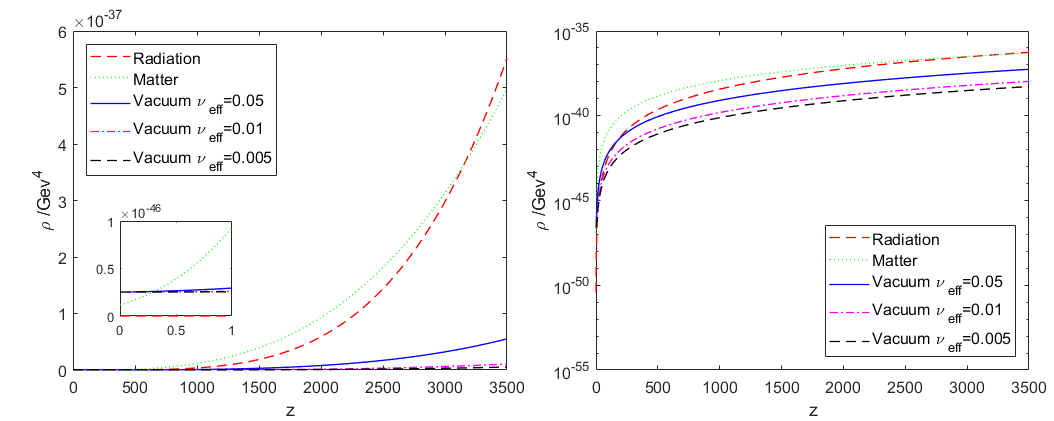}
  \label{fig:sub1}
\end{centering}
\caption{The plot on the left shows the evolution of the different energy densities with the expansion in the canonical  RVM context, Eq.\,\eqref{eq:RVMcanonical}.  The inner window serves to magnify the  low redshift region. The right plot provides a complementary view using a (vertical) logarithmic scale. The VED  exhibits a very mild dynamics up to the  radiation dominated epoch.}
\protect\label{Fig1}
\end{figure}

\section{The evolving VED in the present epoch: the canonical RVM}\label{sec:CanonicalRVM}

What about  VED physics? The measurable difference between the VED values at different epochs  of the cosmic evolution, say $H_0=H(t_0)$ and $H=H(t)$  within our observational range, now follows from the usual RG prescription,  based on choosing the renormalization points near the corresponding values of  the energy scales, in this case $M_0=H_0$  and $M=H$ (hence bringing  them  near the physical state of the FLRW spacetime  at each  epoch). For simplicity, we denote these  VED values as $\rv(H_0)$  and   $\rv(H)$, respectively.  Using \eqref{DiffHH0MM0}, the  leading result can be cast as follows\,\cite{CristianJoan2020,CristianJoan2022}:
\begin{equation}\label{eq:RVMcanonical}
\rv(H)=\rvo+\frac{3\nueff(H)}{8\pi}\,\mpl^2\,(H^2-H_0^2)+{\cal O}(H^4)\,,
\end{equation}
with $\mpl=G_N^{-1/2}$   the usual Planck mass.  As usual, we shall  neglect  the ${\cal O}(H^4)$ terms for all the considerations referring to the current universe (and for that matter for the entire FLRW regime, which is well  away from the early inflationary period). The effective running parameter  $\nueff(H)$ is  a (mildly evolving)  function of $H$ during the FLRW regime and  is given in the  Appendix \ref{sec:AppendixA},   but for the late time universe  it suffices to take it constant, namely $\nueff\equiv\nueff(H_0)$:
\begin{equation}\label{eq:nuandepsilon}
\nueff\simeq\epsilon \ln\frac{m^2}{H_0^2}\,, \ \ \ \  \ \ \ \ \epsilon\equiv\frac{1}{2\pi}\,\left(\xi-\frac{1}{6}\right)\,\frac{m^2}{\mpl^2}\,.
\end{equation}
Both $\epsilon$ and $\nueff$ are  small parameter since $m^2\ll \mpl^2$ for any particle mass.   Clearly,  the dominant contribution  to the VED running stems from the largest masses $m\sim M_X$, presumably  from fields of a typical GUT scale $M_X\sim 10^{16}$ GeV (possibly including a large multiplicity factor)\,\cite{Fossil2008}.

In the expression \eqref{eq:RVMcanonical} we have identified $\rv(H_0)$ with today's VED value, $\rvo$, while  $\rv(H)$ stands for the VED at a nearby point $H$.  Equation \eqref{eq:RVMcanonical} constitutes the canonical form of the RVM\,\cite{JSPRev2013,JoanSolaPhilTransc2022}.  As noted above,  phenomenological analyses of the cosmological data support this scenario and makes it competitive with the standard model with rigid $\CC$-term\,\cite{EPL2021,RVMphenoOlder1,RVMphenoOlder2,AdriaJoan2018,CosmoTeam2022}.  Worth noticing, the RVM passes also successfully  the basic cosmographic tests\,\cite{Mehdi}, which are essentially model-independent.

The phenomenological success of the above RVM formulas seem to effectively support  the fact that  for the FLRW universe  the natural choice of the scale $M$  is indeed  $M=H$. It has the triple virtue of being:  i) theoretically consistent (as shown in the comprehensive works \cite{CristianJoan2022,CristianJoan2020}),  ii)  inspired in the usual practice of ordinary gauge theorizes, as previously noticed,  and  iii) it identifies the proper energy scale (in natural units)  of the FLRW cosmology.   Such a scale setting is clearly adapted for the study of the homogeneous and isotropic universe as a whole, therefore satisfying the Cosmological Principle. However, extending it to smaller scales is a delicate matter, especially if one wants to be free from model-dependent assumptions and still be able to probe cosmological and astrophysical effects at a time. For instance,  in the context of cluster and galactic systems  there are relevant local scale settings (typically associated with the physical dimensions of the involved structures) that allow to explore the possibility of having bags of  inhomogeneous vacuum energy capable of  influencing the processes of gravitational collapse of these structures.  Examples on how to treat these situations within the RVM  have been considered in the past, see e.g.  \cite{Plionis2010,Grande2011,Adria2015,CosmoAstro},  although there are additional  assumptions to be made in these cases which  unavoidably lead, as mentioned,  to  a model-dependent approach, something which we would like to avoid here since it could obscure the interpretation of the  purely cosmological scenarios based on the homogeneous and isotropic FLRW universe.  It is already  significant from our point of view the fact that we can effectively  test the simplest assumption $M=H$ in the pure cosmological context and find it to be fully consistent with the modern cosmological observations and in particular with the large scale structure formation data in the linear regime,  obtaining quality fits that surpass  the performance of the $\CC$CDM in many cases, as shown by the fact that the $H_0$ and $\sigma_8$ tensions can be highly alleviated\,\cite{EPL2021}.

From the foregoing discussion, we learn that QFT in curved spacetime predicts that the VED is a slowly evolving function of the cosmological expansion, and hence the effective  $\Lambda$-term too  (remember that the physical value of  $\Lambda$ is proportional to $\rv$,  not to the parameter  $\rho_\Lambda$ in the action).  We can better appraise the evolution of the VED in a graphical way in Fig.\,\ref{Fig1}.  Parameters  are taken from the best-fit values of\,\cite{PlanckCollab}.  On the left plot of Fig.\,\ref{Fig1} we show the evolution of the matter densities (relativistic and nonrelativistic)  together with the slow evolution of the vacuum density.  On the right plot we depict a logarithmic representation of the various densities such that the differences can be better appreciated, above all  in the case of the VED. The curves are  displayed for different typical  values of $\nueff$.  Despite of the  fact that the VED evolution is very mild, of course, its EoS is nevertheless potentially observable, see Sec.\,\ref{sec:EoS-now}.

\section{Running gravitational coupling}\label{sec:RunningG}

The evolution of the VED preserves the Bianchi identity provided there is an exchange with another dynamical variable\,\cite{JSPRev2013,JoanSolaPhilTransc2022}. If we assume local matter conservation (i.e. no exchange between the vacuum and any of the matter components such as  dust or radiation density, collectively represented by $\rho_{\rm m}$), then the gravitational coupling $G$ must vary with the cosmic expansion to compensate for the VED running.  Let us  consider the late universe, in which we can neglect the ${\cal O}(H^4)$ renormalization effects on the VED.  Using the formalism of \,\,\cite{CristianJoan2022}  we find that the cosmic time evolution  of the VED is connected to that of $G$ as follows:
\begin{equation}\label{eq:NonConserVEDandG}
\dot{\rho}_{\rm vac}+3H\left(\rv+P_{\rm vac}\right)=-\frac{\dot{G}}{G}\,(\rho_m+\rv)=- \frac{\dot{G}}{G}\,\frac{3H^2}{8\pi G}\,,
\end{equation}
where  Friedmann's equation has been called for under the assumption that the higher order gravitational terms do not contribute in the current universe.   As  indicated above,  to trace the  evolution of the VED at the cosmic history time $H(t)$ one can  take the renormalization scale $M$ at this value and we obtain the desired running law for the gravitational coupling as a function of the Hubble rate.  We find\,\cite{CristianJoan2022}
\begin{equation}\label{eq:GNHfinal}
G(H)=\frac{G_N}{1-\epsilon \ln \frac{H^2}{H_0^2}}\,.
\end{equation}
Notice that  $G_N=G(H_0)$ is the current local gravity value  (Newton's `constant'),  usually associated to the inverse Planck mass squared:  $ G(H_0)=G_N=1/m^2_{\rm Pl}$ (in natural units).  The parameter $\epsilon$ in \eqref{eq:GNHfinal} is the same, of course, as the one previously defined in \eqref{eq:nuandepsilon}.   It is apparent  that for $\epsilon=0$ (hence $\nueff=0)$, both  $\rv$ and $G$ cease to be  running quantities since they do not feel the quantum vacuum effects.  But for $\epsilon\neq0$  ($\nueff\neq 0$) there is indeed a dynamical exchange between the two quantities which insures the perfect fulfilment of the Bianchi identity and shows the consistency of the obtained result. One can also determine the explicit form of the running couplings for the gravitational HD terms\,\cite{CristianJoan2022,CristianJoan2020,Ferreiro2019-20}, but the most relevant running laws for our purposes are those for $\rv$ and $G$.  They are both  necessary to compute the vacuum EoS  (see Appendix \ref{sec:AppendixA}  for details).

A final comment may be in order to further illustrate the potential significance of this framework. Testing the evolution of the VED in curved spacetime through the cosmic dependence of the renormalization scale  is a novel feature as compared  to ordinary gauge theories of strong and electroweak interactions in flat spacetime.  Interestingly,  it makes possible to probe the effect of the (cosmic) time-dependence of the running couplings and masses in the particle and nuclear physics world, and hence it may  ultimately provide a possible theoretical explanation \,\cite{FritzschSola,FritzschCalmet}  for the  purported evolution of the fundamental `constants' of Nature, as claimed in some experiments\, \cite{Uzan2015}.  Modern attempts at challenging the stability of the fundamental `constants' can be seen e.g.  in \cite{CalmetExperiments0222} and references therein.

\begin{figure}
\begin{center}
\includegraphics[width=1\textwidth]{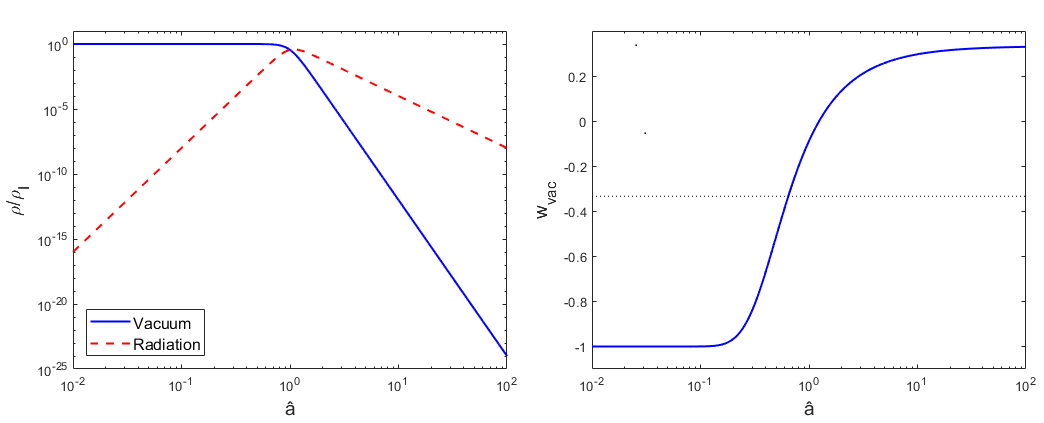}
\end{center}
\caption{Inflationary period.  On the left it is shown the evolution of the energy densities \eqref{rhodensities} of vacuum and relativistic matter before and after  the transition point $a_*\sim 10^{-29}$ (where $\hat{a}\equiv a/a_*$ takes the value $1$)  from inflation to the early radiation epoch (see the text).  The (constant) VED during inflation decays into radiation and the standard FLRW regime starts.  On the right we can see the evolution of the vacuum EoS from $\wv\simeq -1$  up to $\hat{a}=1$.  Once  this point  is left well behind ($\hat{a}\gg 1$), the vacuum evolves into an incipient radiation phase and adopts its EoS:  $\wv\rightarrow 1/3$.}
\protect\label{Fig2}
\end{figure}

\section{EoS of the running  vacuum in the inflationary epoch}\label{sec:EoS-Inflation}

It was recently argued that  inflation could be another consequence of the running vacuum universe\,\cite{CristianJoan2022}.  If so, there would be  no need to introduce explicit, \textit{ad hoc},  inflaton fields in the classical action. In this approach, inflation in the very early universe can be produced by pure quantum effects in QFT in curved spacetime.  To bring about inflation we need (even) powers of $H$ beyond $\sim H^2$, i.e. $H^N\, (N=4,6,...)$.  Inflation then proceeds through a short period where $H=$const.  We call this mechanism  RVM-inflation\cite{CristianJoan2022,CristianJoan2020}, see also \cite{JSPRev2015}.  The needed powers of $H$ emerge from calculating the ZPE up to $6th$ adiabatic order  (not shown in Eq.\,\eqref{Renormalized2}). The $\sim H^4$ ones disappear in the adiabatic subtraction procedure\footnote{See \cite{NickJoan2021} and references therein for a related (stringy) approach.}.  In the present context, therefore,  the $H^6$ terms take over during inflation. Their computation is rather cumbersome\,\cite{CristianJoan2022}, but these terms are finite and do not require renormalization. The final result can be condensed as follows:
\begin{equation}\label{eq:RVMinflation}
\rv^{\rm inf}=\frac{\langle T_{00}^{\delta \phi}\rangle^{\rm 6th}_{\rm Ren}(m)}{a^2}=\frac{\txi}{80\pi^2 m^2}\, H^6+ f(\dot{H}, \ddot{H},\vardot{3}{H}...)\,,
\end{equation}
where we have defined the parameter
 $\txi=\left(\xi-\frac16\right)-\frac{2}{63}-360\left(\xi-\frac16\right)^3$.
The remaining terms are collected in the complicated function $ f(\dot{H}, \ddot{H},\vardot{3}{H}...)$. They  carry along many different combinations of powers of $H$ accompanied  in all cases with time derivatives of $H$, and hence they  all vanish for $H=$const.  This means that a short period where  $H=$const can trigger inflation  from the $\sim H^6/m^2$ term indicated above, where $m\sim M_X\sim 10^{16}$ GeV.
Explicit analytic  solution for the Hubble rate and matter densities during the inflationary epoch  is possible, with the results
\begin{equation}\label{eq:Hinfl}
H(\hat{a})=H_I \left(1+\hat{a}^{8}\right)^{-1/4}
\end{equation}
and
\begin{eqnarray}\label{rhodensities}
\rho_r(\hat{a})=\rI\,{\hat{a}^8}\left(1+\hat{a}^8\right)^{-\frac{3}{2}}\,\,, \ \ \ \ \ \ \ \ \
\rv(\hat{a})={\rI} \left(1+\hat{a}^{8}\right)^{-\frac{3}{2}}\,.
\end{eqnarray}
We see from \eqref{eq:Hinfl}  that in the beginning  the Hubble rate evolves very little around an initial (big) value  $H_I\sim\MPl^{1/2}\, m^{1/2}  \tilde{\xi}^{-1/4}$, namely
$H(\hat{a})\simeq H_I$  for  $0<\hat{a}<1$,  where we have defined $\hat{a}\equiv a/\astar$ and  $\astar$ determines the transition point from the regime of vacuum dominance into that of radiation dominance,  as can be easily inferred  from the density equations \,\eqref{rhodensities}.   The point $\astar$  is estimated to be around $a_* \sim 10^{-29}$ in \,\cite{Yu2020}.  Since $\dot{H}=-2H_I^2 \hat{a}^8/(1+\hat{a}^8)$, we have $|\dot{H}/H^2|\propto\hat{a}^8\ll1$ for $\hat{a} \ll 1$ and we can safely neglect $\dot{H}\approx 0$, and successive derivatives, during inflation.  In  Fig.\,\ref{Fig2} (left) we depict the evolution of the vacuum and radiation densities, where we can see that the vacuum state rapidly decays into radiation,  as it is  also obvious from the two relations in \eqref{rhodensities}.  At the beginning ($a=0$) there is no radiation at all ($\rho_r(0)=0$), whilst the VED  at this point  is maximal, namely   $\rv(0)=\rI\propto\MPl^2H_I^2$, but finite.  This shows in passing that there is no initial singularity in this formulation.  On the other hand, for $\hat{a}\gg 1$ (i.e.  $a\gg\astar$)  it is reassuring to see that we retrieve the standard decaying behavior of  radiation, $\rho_r(a)\sim a^{-4}$.  In the meantime,  the primeval VED decreases very fast and it causes no problem with primordial BBN (big bang nucleosynthesis)  even if $\nueff$ is kept in the radiation epoch  (see next section).  Thus,   RVM-inflation is followed by a standard FLRW radiation epoch.  This type of scenario, which we find here in the context of QFT in curved spacetime,  was assumed phenomenologically in \cite{BLS2013} -- see also the recent  comprehensive study\,\cite{Yu2020}.   We should also clarify that  RVM-inflation is different from Starobinsky's inflation\,\cite{Staro80}, where it is $\dot{H}$ rather than $H$ that remains constant for a short time -- see  \cite{JSPRev2015,NickJoan2021} for a thorough discussion.

Remarkably, during this initial phase we find that the running vacuum behaves as `true' vacuum with equation of state (EoS) $\wv=-1$. Indeed, the vacuum EoS in the early universe follows from computing the corresponding vacuum pressure at that primeval stage up to $6th$ adiabatic order. The result adopts the form:
\begin{equation}\label{eq:VacuumPressureFullsplit}
\Pv(M)=-\rv(M)+f_2(M,\dot{H})+ f_4(M,H,\dot{H},...,\vardot{3}{H})+f_6(\dot{H},...,\vardot{5}{H})+\cdots\,,
\end{equation}
in which the functions $f_2$, $f_4$ and $f_6$ involve  adiabatic contributions of second, fourth and sixth order, respectively,  and all of them carry at least one time derivative of $H$\,\cite{CristianJoan2022}.  Therefore,  all these functions vanish for $H=$const.  ($\hat{a}\ll 1$) and we find  $\Pv=-\rv$  to a very good approximation.  The RVM inflationary period  is thus characterized by the traditional EoS of vacuum, $\wv=-1$. This can be appreciated in Fig.\,\ref{Fig2} (right).

\section{EoS of the running vacuum  in the FLRW regime }\label{sec:EoS-now}

 We have just seen that the vacuum  EoS, $\wv$,  during the inflationary epoch is very close to  $-1$, but the more we  near the radiation epoch the more it  departs from $-1$ and transmutes into $+1/3$, as it can also be clearly seen in Fig.\,\ref{Fig2} (right).  In general, after the inflationary epoch (i.e. for $\hat{a}>1$), quantum effects trigger a fully dynamical behavior of $\wv$ which goes on during the entire conventional FLRW regime. As a result, the vacuum EoS does not remain stuck to the classical value $\wv=-1$ and indeed changes throughout different epochs. Such an evolution can be explicitly derived from the QFT framework of \,\cite{CristianJoan2022,CristianJoan2020}. Some details of the calculation are provided in  the Appendix \ref{sec:AppendixA}, where the precise formula is given.  A sufficiently accurate approximation to the running vacuum EoS during the entire FLRW cosmic stretch  reads as follows:
\begin{equation}\label{eq:EoSChameleon}
\wv(z) = -1+\frac{\nu_{\rm eff}\left(\Omega_{\rm m}^0 (1+z)^3+\frac{4}{3}\Omega_r^0 (1+z)^4\right)}{\Omega_{\rm vac}^0+ \nu_{\rm eff}\left[-1+\Omega_{\rm m}^0 (1+z)^3+\Omega_r^0 (1+z)^4+\Omega_{\rm vac}^0\right]}\,,
\end{equation}
where $\Omega_{\rm vac}^0=\rvo/\rco\simeq 0.7$ is the current vacuum cosmological parameter,  whereas  $\Omega_{\rm m}^0=\rmo/\rco\simeq 0.3 $  and $\Omega_{\rm r}^0=\rho_{\rm r}^0/\rco\sim 10^{-4}$  are the corresponding matter and radiation parts.  Since $|\nueff|\ll 1$ and  $\Omega_{\rm r}^0 \ll \Omega_{\rm m}^0$, it is readily seen that for small $z$ the previous formula boils down to
\begin{equation}\label{eq:EoSDeviation}
\wv(z) \simeq   -1+\nueff \frac{\Omega_{\rm m}^0}{\Omega_{\rm vac}^0}(1+z)^3\ \ \ \ \ \ \ \ \ \ (z< {\cal O}(1))\,,
\end{equation}
thus recovering  the approximate result first advanced in  \cite{CristianJoan2022}.  Here, however,  we have generalized this result into the more complete formula \eqref{eq:EoSChameleon} for the  full FLRW regime  (cf.  Appendix \ref{sec:AppendixA}).


\begin{figure}
\begin{center}
\includegraphics[scale=0.9]{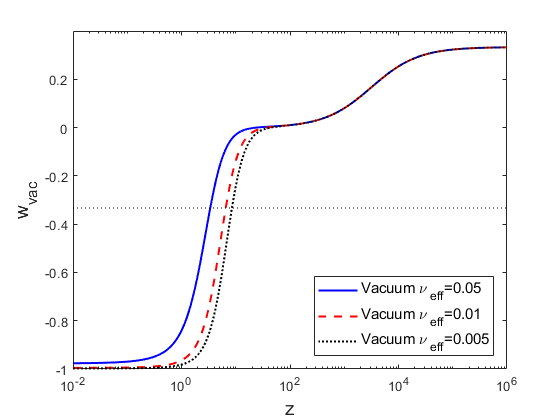}
\end{center}
\caption{Vacuum EoS  for different (positive) values of $\nueff$\,\cite{EPL2021}. Some regimes to be noted: i) $\wv\simeq-1$   for very low redshift, ii)  $-1<\wv<-1/3$, vacuum mimics quintessence for low and intermediate redshift (the horizontal dotted line marks off the DE threshold $\wv=-1/3$), iii) $\wv=0$ plateau, vacuum imitates dust matter, and iv) $\wv=1/3$ plateau, vacuum mimics radiation.  The running vacuum behaves as a cosmic chameleon.}
\protect\label{Fig3}
\end{figure}


The above EoS formulas depend on the crucial coefficient $\nueff$, which we have computed in QFT but it must ultimately be fitted to the cosmological data\,\cite{EPL2021,RVMphenoOlder1,RVMphenoOlder2,AdriaJoan2018,CosmoTeam2022}.  These analyses show that $\nueff\sim 10^{-2}-10^{-3}$  and that  $\nueff>0$ is the preferred sign.

From the foregoing considerations, we find that the running vacuum never has the exact EoS $\wv=-1$ during the FLRW stage,  not even at $z=0$, where
\begin{equation}\label{eq:EoSnow}
\wv(0) \simeq   -1+\nueff \frac{\Omega_{\rm m}^0}{\Omega_{\rm vac}^0}\gtrsim -1 \ \ \ \ \ \ \ \ \ \ (\nueff>0)\,.
\end{equation}
Thus, amazingly,  the RV currently behaves as quintessence\,\footnote{Equation \eqref{eq:EoSDeviation} resembles  previous effective EoS forms for the dynamical VED derived phenomenologically in \cite{SolaStefancic2005,BasilakosSola2014}, although it is different from them since it predicts a quintessence behavior of the RV already at $z=0$,  in contrast to the aforementioned forms which predict a departure of $\wv$ from $-1$ only for $z>0$ but still yield the conventional behavior $\wv=-1$ at $z=0$.}. Such an effective behavior is triggered by the quantum effects and from this point of view there would be no need to introduce \textit{ad hoc}  quintessence fields (nor \textit{ad hoc}  inflatons, as shown in the previous section).


\begin{figure}
\begin{center}
\includegraphics[scale=0.7]{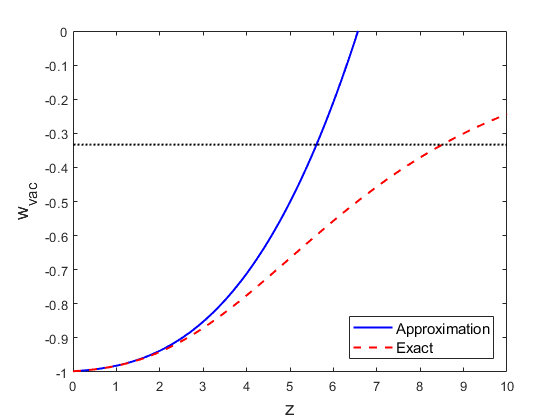}
\end{center}
\caption{Deviation of the approximate vacuum EoS formula \eqref{eq:EoSDeviation}  with respect to the more precise one given by Eq.\,\eqref{eq:EoSChameleon}  for a typical value  $\nueff=0.005$.}
\protect\label{Fig4}
\end{figure}


 In  Fig.\,\ref{Fig3} we provide a detailed plot of the more general formula for the EoS \eqref{eq:EoSChameleon} and for a large window of the FLRW regime spanning from the present time up to high redshift, in fact  covering the entire nonrelativistic matter-dominated (dust) epoch and embracing part of the radiation epoch.  The plot is performed for  different values of $\nueff$ within the typical range obtained in actual fits to the data\,\cite{EPL2021}.  The  approximate EoS \eqref{eq:EoSDeviation}  is only valid for the most recent universe and deviates significantly from the more accurate one \eqref{eq:EoSChameleon}  for intermediate or large values of $z$.  This can be clearly seen in Fig.\,\ref{Fig4}  where the two formulas are plotted on top of each other so as to ease the comparison and to evince the large deviation at higher and higher redshifts.  Notice that the detailed plot of the vacuum EoS in Fig.\,\ref{Fig3}  interpolates  in a numerical way the results that can be directly inferred analytically from Eq.\,\eqref{eq:EoSChameleon} for the different redshift intervals  all the way from the radiation epoch, down to the matter-dominated epoch until reaching the current epoch.  Denoting by  $z_{\rm eq}=\Omega_{\rm m}^0/\Omega_{\rm r}^0-1\simeq 3300$ the equality point between matter and radiation, we find
 \begin{equation}\label{eq:EoSregimes}
\wv(z)=\left\{
\begin{array}{ll}
 \frac13  \ \ \ \  \text{for}\ \ z\gg z_{\rm eq}\ \  \text{with}\ \ \ \Omega_{\rm r}^0 (1+z)\gg\Omega_{\rm m}^0,
 & \text{radiation behavior}\, \, (\nueff\neq0), \\
 & \\
\hspace{0.0cm} 0  \ \ \ \  \text{for}\  \ {\cal O}(1)<z\ll z_{\rm eq}\ \  \text{with}\ \ \ \Omega_{\rm m}^0\gg \Omega_{\rm r}^0 (1+z), &  {\text{dust behavior}} \, \, ( \nueff\neq 0),\\
& \\
-1+\nueff \frac{\Omega_{\rm m}^0}{\Omega_{\rm vac}^0}(1+z)^3\ \ \  \ \text{for}\ \ \  -1<z<{\cal O}(1)\,,
& \text{quintessence behavior}\, \, ( \nueff>0)\,.  \\
\end{array}
\right.
\end{equation}
 As it turns out, the running vacuum EoS follows the EoS of relativistic matter in the radiation-dominated epoch, subsequently the EoS of non-relativistic (dust) matter in the matter-dominated epoch,  the EoS of quintessence at present (for $\nueff>0$) and finally asymptotes to de Sitter era in the future ($z\to-1$).

In the presence of quantum  vacuum effects,   the deceleration parameter $q = -1 -\dot{H}/H^2$ can be easily derived. Using the expression for the quantum corrected $H$ derived in Appendix \ref{sec:AppendixA} up to order ${\cal O}(\nueff)$ and requiring that $q=0$ we find that the transition redshift from deceleration to acceleration becomes slightly shifted with respect to that of the concordance model (aka $\CC$CDM),  as follows:
\begin{equation}
z_t=\left(\frac{2\left(\Omega_{\rm vac}^0-\nueff\right)}{\Omega_{\rm m}^0 (1+\nueff)}\right)^{1/3} - 1\,.
\end{equation}
As expected, the  $\CC$CDM result is recovered for $\nueff=0$. Since, however,  $\nueff$ is small and  $z_t$ cannot be measured with precision yet,  it is not the ideal signature.  What it really acts as a useful signature of the RV  is its effective behavior as quintessence in the low redshift range, as we have seen above.
Indeed, the running vacuum is kind of `chameleonic'. It  behaves as `true' vacuum ($\wv=-1$) only in the very early times when it  triggers inflation. It then  remains silent for eons  (hidden as if being  relativistic or  pressureless matter). Today,  it appears as (dynamical)  dark energy (DE), specifically as  quintessence  ($-1<\wv<-1/3$), cf. Fig.\,\ref{Fig3} .  As a result of this multifaceted  behavior, it may crucially help in solving the $\sigma_8$ and $H_0$ tensions\,\cite{Tensions1,Tensions2,Tensions3,WhitePaper2022} afflicting the  $\CC$CDM model.  In fact, in \cite{EPL2021} it was argued that if there is a  `DE threshold' $z_{*}$ near our time where the DE dynamics of the vacuum gets suddenly activated, this can be extremely helpful for solving the $\sigma_8$ tension within the RVM.  At the same time, it was shown that if the gravitational coupling runs slowly (logarithmically) with the expansion, this can help fixing the $H_0$ tension.  In Fig.\,\ref{Fig3}  we can  see that a continuous (i.e. not abrupt)  DE `threshold' window with low  $z_{*}={\cal O}(1)$  does indeed exist for the RVM, in the sense that for $z<z_{*}$ the vacuum gets progressively activated as DE  ($\wv<-1/3$),  whereas  for $z>z_{*}$ the vacuum EoS  transmutes successively into that of dust matter  and radiation.  There is therefore a tracking of the matter EoS by the vacuum in the RVM framework.

Some of the dynamical properties exhibited by the running vacuum in the current QFT formulation\,\cite{CristianJoan2022,CristianJoan2020} have been longed for  in the past using  \textit{ad hoc} scalar fields in the classical action, see e.g. \cite{PeeblesRatra2003}  and references therein.   In fact, many authors have tried to motivate  a dynamical character of the dark energy (DE)  through cosmological scalar fields (quintessence and the like) since this could help solving the cosmic coincidence problem\,\cite{Steinhardt2003}.   This  can be achieved by picking out the effective potential of the scalar field among those that satisfy the so-called  tracker condition.    In these cases the effective EoS of the scalar field can track matter through the cosmic evolution, see e.g. \cite{Sola:2016hnq} where the tracking feature  is illustrated for the well-known Peebles \& Ratra potential\,\cite{PeeblesRatra1988}.  Here, in contrast, we have shown for the first time (to the best of our knowledge) that the quantum vacuum associated with the quantum fluctuations of the matter fields (in the context of QFT in curved spacetime) has the ability to track the EoS of matter throughout the cosmic evolution and can mimic quintessence in the late universe.  Interestingly enough, this feature is accomplished here by virtue  of the inner dynamics of the quantum vacuum.   In the absence of the quantum  fluctuations of the quantized matter fields,  the vacuum EoS would be stuck at $-1$, as usually assumed. We believe that this remarkable new  ingredient of the QFT formulation of the  RVM (which was entirely absent in the old proposals, see \cite{JSPRev2013} and references therein)  is worth being stressed. In point of fact, it is one of the main results presented in this work.

Finally, let us  recall that in the present RVM framework the dynamics of  vacuum is intertwined with that of the gravitational coupling through a log of the Hubble rate: $G=G(\ln H)$\,\cite{CristianJoan2022}.  This fact together with the mentioned tracking feature (which  is responsible for  the aforementioned existence of a DE `threshold' window at low redshift)  are both present and  they  combine constructively to mitigate the $\sigma_8$ and $H_0$  tensions at a time.  The running vacuum EoS for the current universe \eqref{eq:EoSDeviation}  is actually similar to the EoS of the  effective dark energy (DE) in a Brans-Dicke (BD)  theory in the presence of a cosmological constant, as in fact such theory mimics the RVM -- see Ref.\,\cite{BDCosmoTeam}.   Additionally, the trademark of the  BD framework is indeed the existence of a mildly varying $G$.  In this respect let us  note that recent phenomenological analyses on the viability of different kinds of modified gravity theories have put tight constraints from  BBN  on their parameters, see e.g.  the work \cite{Asimakis2022}.  Basically, any  deviation from standard cosmology modifies the expansion rate and hence modifies the freeze-out temperature of the weak interaction processes which control the neutron abundance at the BBN time.  Thus, since a variation of $G$ and/or of the vacuum (in general of the DE) energy density can modify the expansion rate, a bound ensues for the parameters of the new model. In particular, in the mentioned work \cite{Asimakis2022} an updated BBN bound is put  on the parameter $\nueff$ of the RVM, which is in the ballpark of $10^{-3}$ (being however insensitive to its sign). This updated BBN bound on $\nueff$ turns out to be in accordance with  the typical fitting values  obtained from  the current-era cosmological data  in the last few years, see the various works \cite{EPL2021,RVMphenoOlder1,RVMphenoOlder2}.  In short, the competitive fits to the global cosmological data obtained from the RVM, which  in fact challenge the performance of the $\CC$CDM,  are consistent with the most recent bounds from BBN.

\section{Conclusions}

The main aim of this work has been to study the equation of state (EoS) of the running vacuum within  the theoretical framework recently  expounded in great detail in\,\cite{CristianJoan2022,CristianJoan2020}, in which the vacuum energy density (VED) is computed for a quantized scalar field nonminimally coupled to gravity in the context of QFT in FLRW spacetime. While the running vacuum model (RVM) idea existed since long on semi-qualitative grounds, the  QFT approach of \cite{CristianJoan2022,CristianJoan2020} --- see \cite{JoanSolaPhilTransc2022} for the essentials and a list of references --  puts a more solid theoretical underpinning to the RVM  and leads to new features which had never been explored before.   In fact, on the basis of this formalism  and in contradistinction to the usual assumption $\wv=-1$, we have found that quantum effects make $\wv$ dynamical and trigger a small deviation of if from $-1$.  We have quantified this deviation by explicitly  computing $\wv$ as a function of the cosmological redshift  for the whole FLRW regime.  The result points to potentially significant  phenomenological implications which can be observationally tested.  In the QFT formulation of the RVM, the dynamics of the EoS actually stems  from the dynamical character of the vacuum itself.   Thus, the measured  value  $\rv(H_0)\equiv\rvo$  does not appear in this framework as a  `fundamental constant' but just as the current value of the VED as a slowly evolving dynamical variable.  Because of the unavoidable need of renormalization in QFT, there is no strict cosmological constant conceived as  an everlasting fundamental entity of Nature. Using the subtracting point $M$  as a renormalization group tool  to explore the cosmic evolution at each expansion history time $H(t)$,  we find that  the VED, $\rv(H)$,  is dynamical and evolves with the cosmic expansion.  However,  the time evolution of the VED is so mild that it mimics the behavior of a `cosmological constant' $\CC=8\pi G_N \rv(H)$ for a large stretch of cosmic time around any given epoch $H$.  In fact, the change is only of order $\sim \nueff H^2$, where the small coefficient $\nueff$ is computable from QFT and is responsible for the minute running of the VED ($|\nueff|\ll 1$). Perhaps the most remarkable point of this result is that such a small evolution can be derived  from  first principles, as in fact $\nueff$ is nothing but  the coefficient of the $\beta$-function of the running VED.

 During the FLRW regime, the  dynamical VED is given by Eq.\,\eqref{eq:RVMcanonical}.   Notwithstanding the small quantum effects encoded in the value of $\nueff$,  the RVM carries two important signatures worth being mentioned owing to their possible phenomenological significance.  First of all,  we emphasize again that  its  EoS  is not given by the constant value $\wv=-1$,  which  has been a characteristic  of the classical vacuum;  rather,  it is time evolving and ultimately an explicit function of the redshift: $\wv=\wv(z)$. Second, the EoS dynamics carries a measurable imprint at present since it behaves as quintessence: $\wv(z)\gtrsim -1$.  There are no quintessence fields at all here, of course; the effective quintessence behavior  is just the consequence of the underlying quantum vacuum effects.   Thus, no  classical \textit{ad hoc} fields are  needed to explain the cosmic acceleration within the RVM framework, as it can be accounted for by the fluctuations of the quantized matter fields\,\cite{CristianJoan2020, CristianJoan2022}.

The nontrivial modification of the EoS of the running vacuum with respect to the classical result $\wv=-1$  is a clear sign that a proper renormalization of the quantum matter  effects was mandatory in the study of the QFT vacuum in a curved background. Not only so, it serves as an effective phenomenological  signature to test the RVM.  Unfortunately, for some time the widespread confusion in the literature about cosmological constant, $\CC$, and VED, $\rv$, has prevented to achieve a proper treatment of the renormalization of these quantities in cosmological spacetime.  Perhaps the most pernicious practice has been the reiterated attempts to relate these concepts in the context of flat spacetime calculations, which is meaningless, see \,\cite{JoanSolaPhilTransc2022}. In flat spacetime one can still define the VED, of course,  but it has no relation whatsoever with the cosmological constant.  As indicated in Sec. \ref{sec:VEDMinkowski}, if we speak of $\CC$ as the physically measured value, then its relation with the current $\rv$ is totally straightforward: $\rvo=\CC/(8\pi G_N)$.  However, at a more formal level where these quantities are derived from a gravitational action in curved spacetime and in the presence of quantized matter fields subject to renormalization, then a lot more of care needs to be exercised.  Leaving for the moment quantum gravity considerations for a better  future (viz. for when the quantum treatment of the gravitational field becomes, hopefully,  accessible),  the more pedestrian renormalization of $\rv$ within QFT in curved spacetime proves to be already quite helpful at present. Because of inappropriate renormalization schemes and computational procedures,  the presence of  quartic mass terms $\sim m^4$  proved to be troublesome within the usual methods, but these difficulties might well be overcome in the formulation presented in \,\cite{CristianJoan2022,CristianJoan2020} on which the present study is based. It leads to a  renormalized VED  which is a mildly dynamical quantity evolving with the cosmic expansion. The outcome is that  $\rv=\rv(H)$ is a smooth function of the Hubble rate and its time derivatives without any disruption from $\sim m^4$ effects\,\cite{CristianJoan2022,CristianJoan2020}.
 In the remote past, however, the higher powers of $H$ (predicted in this approach)  became extremely active and may have triggered fast inflation during a short period in which $H\simeq$ const. At present, on the other hand, a new and much placid de Sitter epoch takes over gradually.
 Overall, the running vacuum acts  as a formidable cosmic chameleon: early on, it triggers inflation as `true vacuum' ($\wv=-1$); then it hides behind matter for aeons (even adopting its EoS: $\wv=1/3$ first,  and $\wv=0$  later); and, finally, it reappears  disguised as  quintessence in our days. Only in the remote future it will become  `true vacuum' again.  The running vacuum reveals itself as a time-evolving entity whose EoS  is also dynamical and changes significantly over the cosmic evolution.  Remarkably,  in the late universe plays the role of (dynamical) dark energy and could afford a reasonable  explanation for the cosmic acceleration.

\vspace{0.5cm}

{\bf Acknowledgments}: We are funded by projects  PID2019-105614GB-C21 (MINECO), 2017-SGR-929 (Generalitat de Catalunya) and CEX2019-000918-M (ICCUB).  CMP is also supported  by  fellowship 2019 FI$_{-}$B 00351.  JSP  acknowledges participation in  the COST Association Action CA18108  ``Quantum Gravity Phenomenology in the Multimessenger Approach  (QG-MM)''.

\vspace{0.5cm}

\appendix
\section{Derivation of the running vacuum EoS for the FLRW regime}\label{sec:AppendixA}
\vspace{0.5cm}

Our goal in this appendix is to provide some details about the derivation of the important EoS formula \eqref{eq:EoSChameleon} given in the main text, which is valid for the  post-inflationary epoch, i.e. for the whole FLRW regime.  For this we will be using the approach and formulae from \cite{CristianJoan2022}.  In the latter reference the running vacuum EoS was disclosed as a function of the redshift only within the approximation $z\ll1$, but here we wish to provide a close expression for the EoS as a function of $z$ valid for the entire FLRW epoch. As previously warned, for all the considerations made  during the FLRW regime we will neglect the quantum corrections of order $ {\cal O}(H^4)$ or above, which can only be relevant for the inflationary epoch.  Thus, for the EoS determination during the post-inflationary epoch,  it suffices  to keep  the terms of adiabatic orders $2$ in Eq.\, \eqref{eq:VacuumPressureFullsplit} only. We find
\begin{equation}\label{eq:wvFLRW}
\begin{split}
\wv(H)=\frac{\Pv(H)}{\rv(H)}=-1+\frac{1}{\rho_{\rm vac}(H)} \frac{\left(\xi-\frac{1}{6}\right)}{8\pi^2}\dot{H}m^2\left(1-\ln\frac{m^2}{H^2}\right) + {\cal O}(H^4)
\end{split}
\end{equation}
where  $\rv(H)$ in the denominator of the above formula is given by  Eq.\,\eqref{eq:RVMcanonical}.  The $ {\cal O}(H^4)$ terms are to be neglected hereafter.
We can see from Eq.\,\eqref{eq:wvFLRW} that at leading order the vacuum EoS is  coincident with that of the $\CC$CDM  ($\wv=-1$),  as it could not be otherwise.  Up to second adiabatic order, it reads
\begin{equation}
\wv(H) =-1+\frac{\epsilon \mpl^2}{4\pi\rho_{\rm vac}(H_0)}\dot{H}\left(1-\ln\frac{m^2}{H_0^2}\right)\simeq  -1-\nueff\, \mpl^2\,\frac{\dot{H}}{4\pi\rvo}\,,
\end{equation}
where the small parameter $\epsilon$ is defined by Eq.\,\eqref{eq:nuandepsilon}. We have set $H=H_0$  in the log since the change is extremely slow within long cosmological periods, for example around our time, and used $\ln\frac{m^2}{H_0^2}\gg 1$ in the last step.  This expression is the result at ${\cal O}(\nueff)$ for very low redshift and  coincides with the result already reported in \cite{CristianJoan2022}. Upon using \eqref{eq:nuandepsilon} and the $\CC$CDM form for $\dot{H}$ (which is consistent at this order)  it can be immediately be written in terms of the redshift as indicated  in  Eq.\,\eqref{eq:EoSDeviation}  of the current work.

However, we would like to generalize that formula for a broader redshift range within the FLRW epoch and for this we cannot approximate the denominator of  \eqref{eq:wvFLRW} through the constant $\rvo=\rv(H_0)$ as we did before.  We need to use now  the dynamical form of the VED during the FLRW epoch, i.e. Eq\,\eqref{eq:RVMcanonical}, in which the  parameter $\nueff$  itself is running\,\cite{CristianJoan2022}:
\begin{equation}\label{eq:nueff2}
\nueff(H)\equiv \epsilon\left(-1+\ln \frac{m^2}{H^{2}}-\frac{H_0^2}{H^2-H_0^2}\ln \frac{H^2}{H_0^2}\right)\,.
\end{equation}
Its approximately constant form for $H$ in the late time universe is given by \eqref{eq:nuandepsilon} in the main text.  To find out the vacuum EoS  such that it be valid for any redshift from now up to the initial stages of the radiation-dominated epoch, we have to insert Eq.\,\eqref{eq:nueff2} into the canonical RVM form for the VED, i.e. Eq.\,\eqref{eq:RVMcanonical},  and use the latter in the denominator of the EoS equation\,\eqref{eq:wvFLRW}.  To further proceed we need an explicit form for $H$. For $\nueff$ strictly constant, the RVM can be solved analytically\,\cite{RVMphenoOlder1,RVMphenoOlder2}.  However, the QFT form of the RVM is more complicated since the effective parameter  \eqref{eq:nueff2} is a function of $H$ and then an exact analytical solution is not feasible. Even so, taking into account that  $\nueff(H)$ is a slowly varying function of $H$ and that  $|\epsilon|\ll 1$,  the function $\nueff(H)$ remains always small,  and hence we can obtain a very good approximate solution for the full FLRW regime by expanding the solution in the small parameter $\epsilon$.  In this way we will be able to split the corrected $H^2$ (involving the QFT effects) into the leading  $\CC$CDM part plus ${\cal O}(\epsilon)$ corrections or higher.  The standard or concordance  $\CC$CDM model part of $H^2$ is simply
\begin{equation}\label{eq:H2LCDM}
{H^2_{\Lambda\rm{CDM}}(z)} = H_0^2\left[\Omega_{\rm m}^0 (1+z)^3+\Omega_{\rm r}^0 (1+z)^4+\Omega_{\rm vac}^0\right] \,.
\end{equation}
Now upon inserting  Eq.\,\eqref{eq:RVMcanonical}  into  Friedmann's equation and separating the $\CC$CDM contribution, we find the following result:
\begin{equation}\label{eq:H2Oepsilon}
H^2=\frac{8\pi G(H)}{3}\left(\rho_{\rm m} (z)+\rho_{\rm vac}(H)+\cdots\right)\simeq H^2_{\Lambda CDM}+\epsilon \left(H^2_{\Lambda \rm{CDM}}-H_0^2\right)\left(-1+\ln \frac{m^2}{H_0^2}\right)+{\cal O}(\epsilon^2)\,,
\end{equation}
where the dots in the first equality stand for the neglected  ${\cal O}(H^4)$ corrections to Friedmann's equation in the present universe (the interested reader can find their explicit form  in \cite{CristianJoan2022}).   In the above expression, the term departing from the $\CC$CDM result  has been calculated up to order ${\cal O}(\epsilon)$, but we should remark  that $G(H)$ in \eqref{eq:H2Oepsilon} is given by by  Eq.\,\eqref{eq:GNHfinal} and hence  it had also to be expanded to ${\cal O}(\epsilon)$ so as to obtain the complete ${\cal O}(\epsilon)$ correction indicated in Eq.\,\eqref{eq:H2Oepsilon}.    In a similar way  we find
\begin{equation} \label{DerivativeOfHExpanded}
\dot{H}=\dot{H}_{\Lambda \rm{CDM}}+\epsilon\dot{H}_{\Lambda \rm{CDM}}\left(-1+\ln\frac{m^2}{H_0^2}\right)+{\cal O}(\epsilon^2)\,.
\end{equation}
Finally, introducing the above equations in Eq.\,\eqref{eq:wvFLRW},   we arrive after some calculations at the formula
\begin{equation}\label{eq:EoSepsilon1}
\begin{split}
&\wv(z)\simeq -1+\frac{\nu_{\rm eff}\,\left(1-\frac{\ln E_{\Lambda \rm{CDM}}^2}{\ln \frac{m^2}{H_0^2}}\right)\left(\Omega_{\rm m}^0 (1+z)^3+\frac{4}{3}\Omega_{\rm r}^0 (1+z)^4\right)}{\Omega_{\rm vac}^0+ \nu_{\rm eff}\left[-1+E_{\Lambda \rm{CDM}}^2(z)\left(1-\frac{\ln E_{\Lambda \rm{CDM}}^2(z)}{\ln \frac{m^2}{H_0^2}}\right)\right]}\,,
\end{split}
\end{equation}
in which $E_{\Lambda \rm{CDM}}^2(z)\equiv \frac{H^2_{\Lambda \rm{CDM}}(z)}{H_0^2}$,   with $\nueff$  given by \eqref{eq:nuandepsilon}.  Once more  we have used  $\ln\frac{m^2}{H_0^2}\gg 1$ to simplify the final result.  In practice, it is sufficient to use the even more simplified form
\begin{equation}\label{eq:EoSepsilon2}
\wv(z)= -1+\frac{\nu_{\rm eff}\left(\Omega_{\rm m}^0 (1+z)^3+\frac{4}{3}\Omega_{\rm r}^0 (1+z)^4\right)}{\Omega_{\rm vac}^0+ \nu_{\rm eff}\left[-1+E_{\Lambda \rm{CDM}}^2(z)\right]}\,,
\end{equation}
since $\frac{\ln E_{\Lambda \rm{CDM}}^2}{\ln \frac{m^2}{H_0^2}}\ll 1$ in the entire FLRW regime, as it can be easily checked.
We immediately recognize that the obtained Eq.\,\eqref{eq:EoSepsilon2}  is just our EoS formula  \eqref{eq:EoSChameleon} in the main text (q.e.d.).  It is fully model-independent  as the mass of the scalar particle has been absorbed by the generalized coefficient  $\nueff$ (within the very good approximation used to derive it). Moreover, as indicated in Sec.\,\ref{sec:EoS-now},  for small redshif values  Eq.\,\eqref{eq:EoSepsilon2} trivially reduces to the much simpler form \eqref{eq:EoSDeviation}. Recall that the three distinct qualitative behaviors implied by the running vacuum EoS  during the various epochs of the  FLRW regime are summarized in Eq.\,\eqref{eq:EoSregimes}.

The above  EoS  formula for the  vacuum can still  be further refined  to include the next-to-leading  ${\cal O}(\nueff^2)$ terms.  This  implies  more work since  we need to consistently collect all of  $\epsilon^2$ terms and in particular  also those from expanding up to that order  the running gravitational coupling  \eqref{eq:GNHfinal}.   We shall omit  the details of this  lengthier calculation.  The result stays, however, rather compact and we find that  up to  the next-to-leading order  in $\epsilon$ we have
\begin{equation}\label{eq:H2Oepsilon2}
H^2(z)=H^2_{\Lambda \rm{CDM}}+\epsilon \left(H^2_{\Lambda \rm{CDM}}(z)-H_0^2\right)\left(-1+\ln \frac{m^2}{H_0^2}\right)+\epsilon^2\left(H^2_{\Lambda \rm{CDM}}(z)-H_0^2\right)\left(-1+\ln \frac{m^2}{H_0^2}\right)^2
\end{equation}
or
\begin{equation}\label{eq:H2Oepsilon2b}
E^2(z)\equiv \frac{H^2(z)}{H_0^2}\simeq \, E_{\Lambda \rm{CDM}}^2(z)+\nueff \left(E_{\Lambda \rm{CDM}}^2(z)-1\right) +\nueff^2 \left(E_{\Lambda \rm{CDM}}^2(z)-1\right)
\end{equation}

and
\begin{equation} \label{DerivativeOfHOepsilon2}
\begin{split}
\dot{H}=\dot{H}_{\Lambda \rm{CDM}}+\epsilon\dot{H}_{\Lambda \rm{CDM}}\left(-1+\ln\frac{m^2}{H_0^2}\right)+\epsilon^2\dot{H}_{\Lambda \rm{CDM}}\left(-1+\ln\frac{m^2}{H_0^2}\right)^2
\simeq  \dot{H}_{\Lambda \rm{CDM}}+\nueff\dot{H}_{\Lambda \rm{CDM}}+\nueff^2\dot{H}_{\Lambda \rm{CDM}}\,.
\end{split}
\end{equation}
These expressions  obviously extend the previous ones up to ${\cal O}(\epsilon^2)$.  We can use them to compute the EoS at this order. Once more we see that the expansion in $\epsilon$ is such that at leading order  it can be expressed as an expansion in $\nueff$. The final result for the EoS to ${\cal O}(\nueff^2)$ takes on the form in   Eq.\eqref{eq:EoSepsilon1} with only the replacement $\nueff\to \nueff(1+\nueff)$ in the parameter $\nueff$ of its numerator.  Thus, since   $0<\nueff\ll 1$,  the next-to-leading ${\cal O}(\nueff^2)$ terms  obviously imply a tiny correction to the ${\cal O}(\nueff)$ formula, which in practice can be neglected.

We remark  that the model at this point is solved. Indeed, from Eq.\,\eqref{eq:H2Oepsilon2b} the quantum correction to  the ordinary $\CC$CDM parameter  $\Omega_{\rm vac}^0$  can be expressed directly in terms of the redshift  as follows:
\begin{equation}\label{eq:rvz}
\Omega_{\rm vac}(z)\simeq \, \Omega_{\rm vac}^0+\nueff \left(E_{\Lambda \rm{CDM}}^2(z)-1\right) +\nueff^2 \left(E_{\Lambda \rm{CDM}}^2(z)-1\right)\,.
\end{equation}
Obviously $\Omega_{\rm vac}(z=0)= \Omega_{\rm vac}^0$ is satisfied, as it should be.  To within  ${\cal O}(\nueff)$ this expression is similar to the one found in previous calculations based on the phenomenological RVM, see e.g. \,\cite{RVMphenoOlder1,RVMphenoOlder2},  except that here we have derived the fundamental RVM formulas, including the running vacuum EoS,  from QFT in curved spacetime within the  framework recently  put forward in \,\cite{CristianJoan2022,CristianJoan2020}.  The above equation can be written to ${\cal O}(\nueff)$  in terms of the vacuum energy density itself   as follows:
\begin{equation}\label{eq:VEDz}
\rv(z)\simeq  \rvo+\nueff\,\rco \left(E_{\Lambda \rm{CDM}}^2(z)-1\right)\,,
\end{equation}
where $\rco=3H_0^2/(8\pi G_N)$ is the current critical density.   This expression has been used for the VED plots in Fig.\,\ref{Fig1}.

\vspace{1cm}

\end{document}